\newcommand{\bk}{\textbf{k}}
\newcommand{\bq}{\textbf{q}}

\documentclass[aps,prb,twocolumn,groupedaddress]{revtex4}
\usepackage{graphicx}
\usepackage{float}
\usepackage{bm}
\usepackage{color}

\begin{document}


\title{Material and doping dependence of the nodal and anti-nodal dispersion 
renormalizations in single- and multi-layer cuprates}


\author{S. Johnston$^{1,3}$}
\author{W. S. Lee$^{2,3}$}
\author{Y. Chen$^{3,4}$}
\author{E. A. Nowadnick$^{3,5}$}
\author{B. Moritz$^{3}$}
\author{Z.-X. Shen$^{2,3,5,6}$}
\author{T. P. Devereaux$^{2,3}$}
\affiliation{$^1$Department of Physics and Astronomy, University of Waterloo, Waterloo, ON N2L 3G1, Canada}
\affiliation{$^2$Geballe Laboratory for Advanced Materials, Stanford University, Stanford, California 94305, USA}
\affiliation{$^3$Stanford Institute for Materials and Energy Science, 
SLAC National Accelerator Laboratory and Stanford University, Stanford, CA 94305, USA}
\affiliation{$^4$Advanced Light Source, Lawrence Berkeley National Laboratory, Berkeley, California 94720, USA}
\affiliation{$^5$Department of Physics, Stanford University, CA 94305, USA}
\affiliation{$^6$Department of Applied Physics, Stanford University, CA 94305, USA}
\begin{abstract}
In this paper we present a review of bosonic renormalization effects on electronic carriers observed 
from angle-resolved photoemission spectra in the cuprates.  We specifically discuss the viewpoint that 
these renormalizations represent coupling of the electrons to the lattice, and review how the wide range 
of materials dependence, such as the number of CuO$_2$ layers, and the doping dependence can be 
straightforwardly understood as arising due to novel electron-phonon coupling.
\end{abstract}

\date{\today}
\maketitle

\section{Introduction}
The discovery of a ``kink" in the nodal ((0,0) - ($\pi$,$\pi$)) dispersion of the high-T$_c$ cuprates,  
and band renormalizations, in the form of a peak-dip-hump structure in the anti-nodal (0,$\pi$) - 
($\pi$,$\pi$) dispersion \cite{BogdanovPRL2000, JohnsonPRL2001, LanzaraNature2001, KaminskiPRL2001, 
CukPSS, DessauPRL1991, CukPRL2004, GromkoPRB2003, KimPRL2003, SatoPRL2003, NormanPRL1997, BorisenkoPRL2006, 
BorisenkoPRL2006-2, KordyukPRL2006, MeevasanaPRL2006, ZhouNature2005, DahmNature2009}, 
have attracted considerable attention in 
recent years.  These band renormalizations are interpreted as due to electron-boson coupling, and 
it is believed that understanding their origin will provide information about the underlying pairing 
mechanism in these materials. There is still considerable debate as to the identity of the 
responsible bosonic mode\cite{mouse}. 
 
One proposal is coupling to an electronic mode associated with the collective mode 
found in neutron scattering near momentum transfers ($\pi$,$\pi$), the so called 
magnetic resonance mode \cite{KaminskiPRL2001, KimPRL2003, NormanPRL1997, BorisenkoPRL2006}.   
Additionally, renormalizations could also be due to coupling of electrons to damped magnons, 
which have less well-defined momentum structure. However, the strength of the 
coupling of magnons to electrons is under debate\cite{Kivelson, Abanov}. 
For example, quantitative comparisons of ARPES and 
neutron measurements on YBa$_2$Cu$_3$O$_{6.6}$ (YBCO) have been made, and the overall strength 
of the coupling inferred from the data was indicated to be of sufficient strength to give 
rise to superconductivity\cite{DahmNature2009}.  
We also remark that a quantitative comparison between the neutron scattering 
and ARPES measurements reported in Ref. \onlinecite{DahmNature2009} can be complicated by the 
polar surface of cleaved YBCO resulting in surface reconstruction.  This can produce 
significant differences between the bulk and surface layers of this material.  
Moreover, a systematic examination of the effects of doping, 
material class, and temperature have not been thoroughly explored. In fact, given that 
the spin continuum arises from the CuO$_2$ plane, one might expect the coupling to electrons 
to be relatively material-class independent.  On the other hand, it is well-known that a 
neutron resonance displays a material dependence, appearing at larger energies for larger T$_c$ 
materials including both the single and multi-layer cuprates.  This opens the possibility of 
linking the neutron resonance with ARPES renormalizations via a material-dependent study.   

An alternative proposal is coupling to a spectrum of oxygen vibrational phonon modes 
\cite{LanzaraNature2001, CukPRL2004, tpdPRL2004, LeePRB2008},   
specifically, the $c$-axis out-of-phase bond-buckling oxygen vibration or B$_{1g}$ mode 
($\Omega \sim 35-45$ meV) and the in-plane bond-stretching oxygen mode ($\Omega \sim 70-80$ meV).  
This proposal has been able to account for many experimental observations including the 
anisotropy of the observed renormalizations\cite{tpdPRL2004}, fine structure in the form of subkinks  
observed in the temperature dependence of the self-energy in 
Bi$_2$Sr$_2$CaCu$_2$O$_{8+\delta}$\cite{LeePRB2008} (Bi-2212),  
and doping dependent changes in self-energy \cite{MeevasanaPRL2006, MeevasanaPRB2006}.  
This interpretation is further supported by recent ARPES experiments that have measured an 
$^{16}O$ $\rightarrow$ $^{18}O$ isotope shift in the nodal kink position\cite{IwasawaPRL2008}. 
Finally, analogous features have been observed in tunneling spectra which have also been 
interpreted as coupling to a bosonic mode \cite{LeeNature2006, Fischer, ZhuPRL2006, ZhuPRB2006}, 
and possibly the same mode responsible for the ARPES observed kink.  Ref.  
\onlinecite{LeeNature2006} reported an $^{18}$O isotope shift of the feature in a Bi-2212 
sample indicating a phonon origin.

In this work, we present a review of how these two scenarios can be differentiated 
by studying the material, doping and temperature dependence of the band renormalizations. 
Our focus is on how $c$-axis phonons provide a material dependence to the ARPES kinks due to 
the phonon's sensitivity to local symmetry and the environment surrounding the CuO$_2$ plane. 
We consider doping dependent changes to the renormalization in Bi-2212 as well as 
the dependence of the renormalization within the Bi and Tl families as the number of CuO$_2$ layers is varied.  
We will also discuss some recent ARPES results on the $n = 4$ layer system Ba$_2$Ca$_3$Cu$_4$O$_8$F$_2$ (F0234)
\cite{ChenPRL2009}.  In this system, the inner and outer layers occupy different crystal environments 
resulting in differing Madelung energies associated with each plane in the  
undoped compound.  This difference drives inequivalent dopings between the two sets of layers, 
with one set $n$-type and the other $p$-type.  The inequivalent doping in each plane 
generates further symmetry breaking in the layers and the el-ph coupling is expected to differ.  
Indeed, Ref. \onlinecite{ChenPRL2009} observes stronger kink features in the plane associated with the 
outer ($n$-type) layer of the material.  Here, we will discuss these observations in the context of 
the el-ph coupling scenario.  

The organization of this paper is follows.  In section II we discuss the role of symmetry breaking 
in producing el-ph coupling to $c$-axis phonons and how such coupling is expected to vary with 
the crystal structure of the high-T$_c$ cuprates.  In section III we present ARPES data for various 
multi-layer cuprates in order to examine how the now the band renormalizations vary with the number of 
CuO$_2$ layers, $n$.  Energy distribution curves in the anti-nodal 
region of the Bi and Tl-families for $n=1-3$ and MDC dispersions for the $p$-bands of the 
F-family with $n=4-5$ are presented.  Of particular interest are the results for the single layer Tl cuprate 
Tl-2201.  Here, we show that Tl-2201 does not resolve the typical peak-dip-hump structures in the antinodal 
region despite the fact that the spin resonance mode exists in this system\cite{LeePRL2009}.      
In section IV the doping dependence of the nodal and antinodal 
dispersions for Bi-2212 are presented. The renormalizations in each region behave differently as 
the samples are overdoped, pointing to presence of multiple bosonic modes.  In section V we 
present a theoretical basis for understanding the self-doping phenomena in the parent compound 
F0234 and discuss what implications this process has on the coupling to $c$-axis phonons.  
Finally, in section VI, we conclude with a brief summary and some additional remarks. 

\section{Electron-phonon coupling in Multi-layer Cuprates}
In this section we discuss how $c$-axis phonons can be sensitive to the material environment off the CuO$_2$ planes. 
Electron-phonon (el-ph) coupling to the in-plane bond-stretching modes are of a deformation type and the 
strength of the coupling depends on the Cu-O bond distance.  Since this distance is relatively constant 
in all cuprates, coupling to these modes is relatively materials independent.   This is not true for 
coupling to the $c$-axis bond buckling modes.  Since our goal is to explore the materials dependence of the 
band renormalizations, we will focus our attention to the $c$-axis modes.  El-ph coupling to $c$-axis phonons can arise to 
first order from modulations in the on-site energy due to atomic site oscillation through local 
crystal fields.  For the CuO$_2$ planes, such crystal fields arise due to asymmetries in the 
environment surrounding the plane \cite{tpdPRB1995, TpdSSC1998,tpdPRL2004}  
and the degree of symmetry breaking varies from material to material with the chemical environment 
(number of layers and doping from the ideal stoichiometric compound).  

In an ideally undoped single layer cuprate the structure above and below the planes is identical and little symmetry 
breaking occurs across the plane.  The coupling to the planar $c$-axis modes is therefore expected 
to be weak in the parent compound.  In the undoped multi-layer systems symmetry breaking across the 
outer planes can be large leading to a sizeable coupling in these layers.  A second pathway for symmetry breaking 
occurs as the parent compounds are doped by introducing substitutional or interstitial dopant atoms 
in the charge reservoir area off the CuO$_2$ planes. 
These dopants donate charge to the CuO$_2$ plane, and cast off-plane electric fields $E_z$ that cannot be 
effectively screened by the in-plane carriers\cite{JohnstonEPL2009}.  As a result  
they form charge impurities that further break symmetry across the CuO$_2$ planes.  Through this mechanism, 
coupling to $c$-axis phonons can occur in single layer systems where they are normally forbidden by 
symmetry.  

\begin{figure}
 \includegraphics[width=\columnwidth]{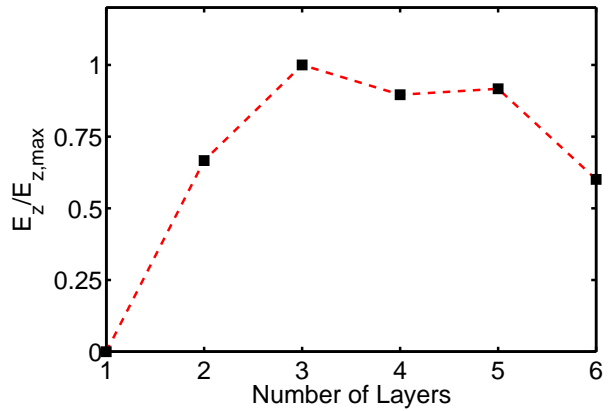}
 \caption{\label{Fig:Efields} The local crystal field strength at the planar oxygen site of the 
 outermost CuO$_2$ plane of the Hg-family of cuprates.  All results have been 
 normalized by the maximum field which occurs for the $n=3$ layer system.}
\end{figure}

To quantify the effect of the crystal environment in the parent systems, we 
calculate the local crystal field at the outermost CuO$_2$ layer for the 
HgBa$_2$Ca$_n$Cu$_n$O$_{2(n+1)+\delta}$ (n = 1-6) family of cuprates.  To do this, we use 
experimental structural data \cite{Hg} and assume an ionic point charge model with formal 
valences assigned to each atom.   The Ewald summation method \cite{Ewald} is then used to perform 
the electrostatic sums for the electric field strength.  The results are shown in Fig. \ref{Fig:Efields}. 

In the single layer compound the local field is zero since the CuO$_2$ layer coincides with the 
plane of mirror symmetry and we have not considered dopings away from pure stoichiometry.  
In the double- and triple-layer compounds the value of the field rises 
taking a maximum for $n = 3$.  After this, as the number of layers further increases, 
there is a general trend of decreasing field values.  
(We note here that the structural data for the $n = 5$ 
and $n = 6$ compounds had a large degree of error, presumably from the difficulty in sample growth.)  
This decrease is due to a reduction in the degree of symmetry breaking across the outer plane once 
the number of layers is increased beyond $n = 3$\cite{Johnston}.  

In the limit of the infinite layer system CaCuO$_2$ the environment is symmetric around every 
plane.  In this case the electrostatic contribution to the coupling is identically zero however 
steric forces introduce buckling to the plane which creates a new pathway for el-ph coupling of 
a deformation type\cite{JLTP}.  However, in terms of material dependence, such steric forces are present 
in all CuO$_2$ systems and therefore do not contribute to differences between materials. 
 
The strength of the coupling to $c$-axis modes, such as the B$_{1g}$ mode, scales as $\lambda \propto E_z^2$.
Furthermore, due to the oxygen charge transfer form factos, the B$_{1g}$ mode couples most strongly to 
antinodal electrons\cite{tpdPRL2004}.  Therefore the behavior of the antinodal renormalization is expected 
to have a dramatic materials dependence if it is due to the B$_{1g}$ phonon.  
This layer dependence is expected to be different for coupling to the spin-resonance mode.  Therefore, by examining the 
change in coupling strength as one moves up in the number of layers provides a pathway for distinguishing 
between these two scenarios.   

\section{Layer Dependence}
In the previous section we discussed how the material dependence of the renormalizations is expected to 
arise in the various families of cuprates and how this coupling is expected to differ in the phonon and 
spin resonance proposals.  We now wish to review the 
available ARPES data in light of the theoretical considerations of the previous section.  
Here, our focus is on the observed changes in the renormalizations as the number of layers within the 
Bi- and Tl-families. 

Single crystals of nearly optimally doped Tl$_2$Ba$_2$CaCu$_2$O$_8$ (Tl-2212), TlBa$_2$Ca$_2$Cu$_3$O$_9$ 
(Tl-1223) and slightly overdoped Tl$_2$Ba$_2$CuO$_6$ (Tl-2201) were grown using the flux method.  As-grown 
Tl2212 (T$_c$ = 107 K) and Tl-1223 (T$_c$ 123 K) crystals were chosen for the ARPES measurements.  
Tl-2201 crystals used in our measurement were prepared by annealing the as-grown crystal (T$_c$ $\sim$ 30 K) 
under a nitrogen flow at a temperature of 500$^\circ$C, yielding a T$_c$ of 80 K.  The data were collected 
using a Scienta R4000 photoelectron spectrometer.  Measurements were performed at the SSRL beam line 5-4 
using 28 eV photons and at the Advanced Light Source beam line 10.0.1 using 50 eV photons.  The energy 
resolution was set at 15-20 meV for the Tl data presented in this work.  Samples were cleaved and measured 
in ultrahigh vacuum ($<4\times10^{-11}$ Torr.) to maintain a clean surface. 
Detailed ARPES results on these compounds have been reported in Ref. \onlinecite{LeePRL2009}. 

\begin{figure}[t]
 \vskip 0.5cm
 \includegraphics[width=\columnwidth]{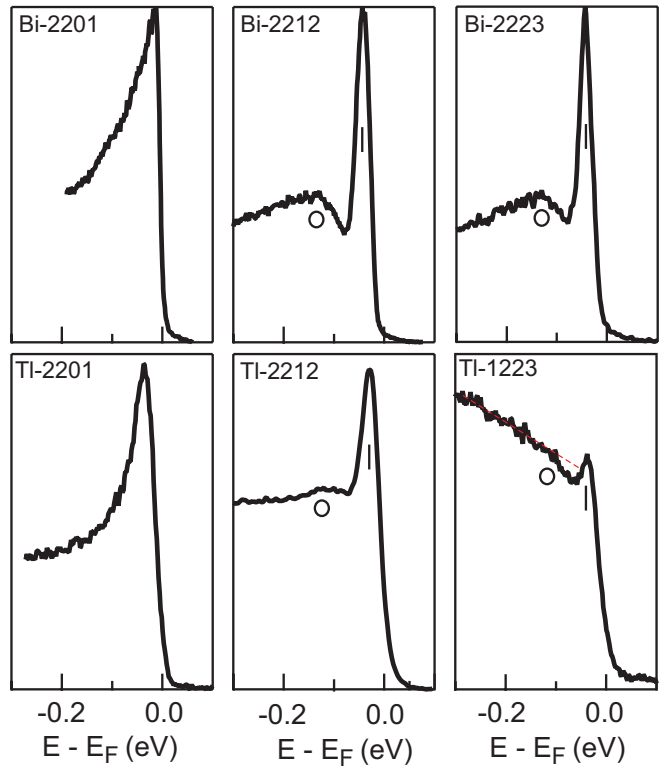}
 \caption{\label{Fig:Layers} 
 Representative EDCs near the antinodal region of the Bi- and Tl-families of cuprates, including single 
 layer (Bi-2201 and Tl-2201), bi-layer (Bi-2212 and Tl-2212), and tri-layer (Bi-2223 and Tl-1223) 
 compounds.  The high background in the data of Tl-1223 is probably due to the absence of a natural cleaving 
 plane in the crystal structure.  Nevertheless, a peak-dip-hump structure in the spectrum can still 
 be discerned.  The red dashed line is a guide-to-the-eye to make the ``hump" more discernible.   
 }
\end{figure}

Although the dispersion kink along the nodal direction has been found universally in high-T$_c$ cuprates 
\cite{LanzaraNature2001}, the momentum dependence of this renormalization feature, when moving away from the  
nodal direction, exhibits a material dependence\cite{LeePRL2009,LeeBook}.  It has  
been confirmed recently that there is dependence on the number of CuO$_2$ planes in the unit 
cell in the Bi-family and, most recently, in the Tl-family of cuprates\cite{LeePRL2009}. In the multi-layer 
compounds, the ``kink" becomes more dramatic and eventually breaks the band dispersion into two 
branches: one branch with a sharp peak and another branch with a broader hump structure \cite{LeePRL2009, LeeBook, 
KaminskiPRL2001, SatoPRL2003, CukPRL2004}.  The two branches asymptotically approach one another at a 
characteristic energy scale of 70 meV and coincide with the dominant energy scale of the kink along 
the nodal dispersion for nearly optimally-doped cuprates.  This separation of the band dispersion 
becomes most prominent near the antinodal region and results in the famous peak-dip-hump 
structure\cite{NormanPRL1997} in the energy distribution curves (EDCs), as shown in Fig. \ref{Fig:Layers}. 

The momentum dependence of the kink is quite different in the single layer compound, where the dispersion kink 
becomes less prominent moving away from the node.  In addition, the band dispersion retains  
a single branch with no separation observed, unlike the case of the multi-layer compounds 
\cite{GrafPRL2008, WeiPRL2008, Footnote}.  As a result, no apparent peak-dip-hump structure can be seen 
in the EDCs near the antinodal region for the single layer compounds Fig. \ref{Fig:Layers}\cite{Footnote}.

In summary, the layer dependent renormalization near the antinodal region is due most likely 
to electrons coupled to a sharp bosonic mode, whose origin is strictly constrained by the 
number of layers in the material.  This mode is either absent, or has a negligible coupling 
to the electrons, in single layer compounds, but exhibits prominent coupling in the multi-layer compounds.  
The spin resonance mode does exist in some single layer cuprates\cite{HeScience2002} (notably Tl-2201 
by not La$_2$CuO$_{4+\delta}$).  Therefore, one can conclude that the spin resonance 
mode is an unlikely candidate for the mode responsible for the renormalizations in the antinodal region.  
On the other hand, coupling to $c$-axis phonons can exhibit a very different coupling in 
single- and multi-layer compounds.  
As we have discussed, the B$_{1g}$ phonon couples strongly to the electrons  
in multi-layer compounds and weakly to electrons in single layer compounds.  This mode can also 
reproduce the observed anisotropic momentum dependence of the renormalization in bi-layer 
Bi-2212\cite{tpdPRL2004}.  We also note that the form for the B$_{1g}$ coupling is attractive 
in the $d$-wave pairing channel\cite{tpdPRB1995}, which could be one factor enhancing 
T$_c$ in the multi-layer systems.   

We now consider another case, that of the F-family of cuprates with $n$ CuO$_2$ layers, $n=3-5$.  
The single crystalline samples were grown by the flux method under high pressure\cite{Structure}.
ARPES measurements on the F-family were performed at beamline 10.0.1 of the Advance Light Source (ALS) 
at Lawrence Berkeley National Laboratory.  The measurement pressure was kept $< 4\times10^{11}$ Torr 
at all time and data were recorded by Scienta R4000 Analyzers at 15K sample temperature.  The total 
convolved energy and angle resolution were 16 meV and 0.2$^\circ$ respectively for photoelectrons generated by 
55 eV photons.   

In Fig. \ref{Fig:Fl_MDC} we present MDC derived dispersions for 
$p$-type bands of the three (F0223), four (F0234) and five (F0245) layer F-based cuprates.  
In all three cases the dispersions show clear kinks, but at a larger energy 
scale in the four- and five-layer materials.  In a $d$-wave superconductor coupled to an Einstein mode, 
the energy scale of the kink occurs at $\Omega + \Delta_0$ where $\Omega$ is the energy of the mode 
and $\Delta_0$ is the maximum value of the superconducting gap.  This shift in energy scale is a reflection 
of the changes in the superconducting gap size as $n$ is varied from 3 to 5.   
In order to quantify the strength of the kink, slopes are extracted from the dispersion above and below the kink 
position, $d\epsilon/dk|_{>}$ and $d\epsilon/dk|_{<}$.  An estimate for the relative 
coupling stengths $\lambda^\prime$ is then given by: 

\begin{figure}[t]
 \includegraphics[width=\columnwidth]{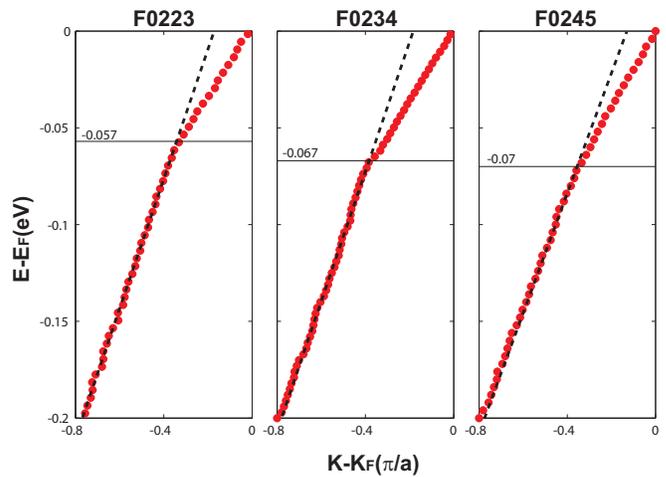}
 \caption{\label{Fig:Fl_MDC} MDC derived dispersions along the nodal direction (0,0) - ($\pi$,$\pi$) 
 of the $p$-type band in the 3-layer (F0223), 4-layer (F0234) and 5-layer (F0245) F-family of 
 cuprates. }
\end{figure}

\begin{equation}
\frac{d\epsilon}{dk}\bigg|_> = (1 + \lambda^\prime)\frac{d\epsilon}{dk}\bigg|_{<}.
\end{equation}

This procedure produces $\lambda^\prime = 0.89$, $0.75$ and $0.49$ for $n = 3$, $4$ and $5$ 
respectively.  This trend is easily understood in the phonon mode scenario where the dominant mode 
in the superconducting state is the B$_{1g}$ mode for which the coupling strength is proportional to the 
local crystal field.  The observed decrease in coupling strength can be easily understood from the 
expected local field strengths in the outer most layers (recall Fig. \ref{Fig:Efields}), which decreases 
for $n > 3$.  

\section{Energy Scales and doping dependence}
In the el-ph coupling picture the carriers couple to a spectrum of bosonic modes and we have already 
seen how the $c$-axis modes can produce a materials dependence of the renormalizations.  It is important to 
note that the coupling to each of these modes is highly anisotropic with the in-plane bond-stretching (breathing) 
mode dominating in the nodal region and the out-of-plane bond-buckling (B$_{1g}$) mode dominating in 
the anti-nodal region.  Since these modes have different frequencies  one would naturally expect the 
multiple energy scales to manifest in the experimental data.  Indeed, evidence for multiple energy scales has 
been found both in the temperature dependence \cite{CukPRL2004, tpdPRL2004, LeePRB2008} as well as the 
doping dependence of Bi-2212 \cite{GromkoPRB2003, MeevasanaPRB2006, MeevasanaPRL2006}.   In this section we 
revisit the doping dependence of the nodal and antinodal renormalizations, highlighting the different behavior 
in each region of the Brillouin zone and discuss how this dichotomy further supports the el-ph scenario.  

High quality single crystals of optimally doped Bi$_2$Sr$_2$Ca$_{0.92}$Y$_{0.08}$Cu$_2$O$_{8+\delta}$ 
(Bi2212 OP, T$_c$ = 96 K) were grown by the floating zone method.  The overdoped crystals with T$_c$ = 88 K 
were prepared by post annealing the optimally doped Bi2212 crystal under oxygen flow at a temperature 
of 400$^\circ$C.  The overdoped sample with T$_c$ = 65K is a derivative of the Bi2212 family with lead 
doped into the crystal to achieve such an overdoped configuration.  The data were collected by using He I 
light (21.2 eV) from a monochromated and modified Gammadata HE Lamp with a Scienta-2002 analyzer and in 
SSRL beamline 5-4 using 19 eV photons with a Scienta-200 analyzer.  The energy resolution is $\sim$10 meV 
and angular resolution $\sim$0.35$^\circ$.  The samples were cleaved and measured in ultra high vacuum 
($< 4\times10^{-11}$ Torr.) to maintain a clean surface.  

\begin{figure}[t]
 \includegraphics[width=\columnwidth]{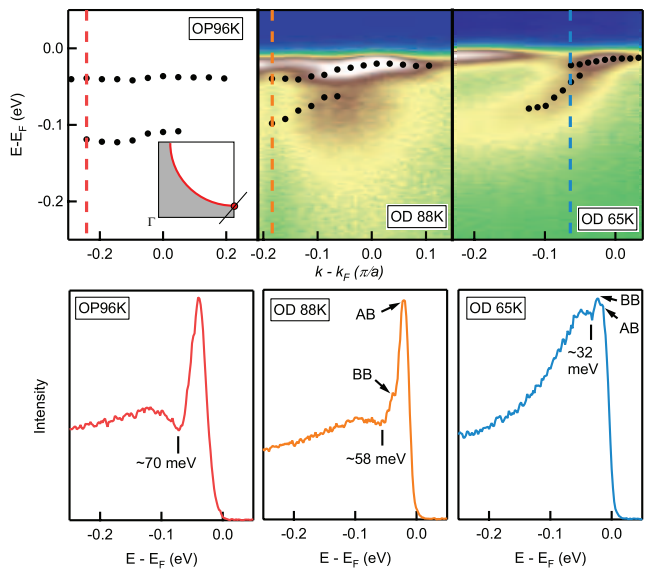}
 \caption{\label{Fig:ScaleAntinode} The doping dependence of the antinodal spectrum of Bi2212 taken in the 
 superconducting state (10K).  Shown in the upper row are the false color plots of the spectra taken along 
 the indicated cut direction (inset).  The black dots are the peak and hump positions of the bonding 
 band seen in the EDCs. Shown in the lower row are the EDCs along the dashed line indicated in the false 
 color plots. The symbols ``AB" and ``BB" represent the antibonding and binding bands while the numbers 
 are the energy position of the dip of the EDC.}
\end{figure}

In Fig. \ref{Fig:ScaleAntinode} ARPES data taken along a cut 
in the antinodal region illustrate this effect.  The upper panels show the measured spectral function along 
the cut while the lower panels show $A(\bk,\omega)$ at a fixed k-point as indicated by the dashed lines.   

Near (0,$\pi$), the energy of the dip feature is the best measure of the energy scale of the mode responsible 
for the renormalization \cite{tpdPRL2004, SandvikPRB2004}.  
For the optimally doped sample (OP96K), shown in the first column of Fig. \ref{Fig:ScaleAntinode}, 
the dip position is clearly located at $\omega \sim 70$ meV.  This can be seen in both the false 
color plot and the EDC cut.    
For moderate overdoping (OD88K), the energy of the dip is lowered to $\sim 58$ meV while for heavily 
overdoped (OD65K) the energy is lowered further to $\sim 32$ meV.  In both overdoped cases, 
contributions from the bonding- (BB) and anti-bonding (AB) bands contribute to the quasiparticle 
peak at the Fermi level.  In the OD65K case, the contribution from the AB makes an exact determination 
of the dip position difficult and the estimate of $\sim 32$ meV should be considered a lower 
bound.   

\begin{figure}[t]
 \includegraphics[width=\columnwidth]{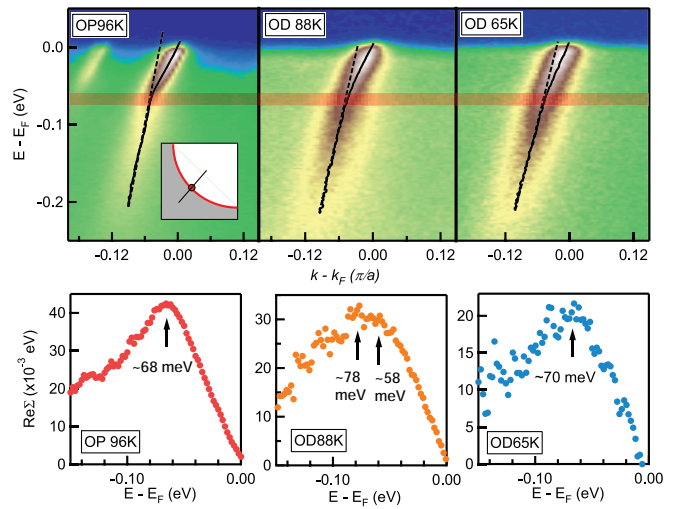}
 \caption{\label{Fig:ScaleNode} The doping dependence of the nodal spectrum at a temperature well 
 below T$_c$ (10 K). Shown in the upper row are the false color plots of the spectra taken along the 
 cut as indicated in the inset.  The black curves are band dispersion obtained by fitting MDCs to 
 Lorentizan functions.  The apparent kink position in the dispersion are marked by the yellow shaded 
 area, which appears to be approximately the same for all three dopings.  The dashed lines serve as a 
 guide-to-the-eye for visualizing the apparent kink in the dispersion. Shown in the lower row are the 
 real part of the self-energy extracted from subtracting the band dispersion from a linear bare band. 
 The arrows indicate the positions of fine structure in the extracted Re$\Sigma$.}
\end{figure}

Turning now to the nodal region, we find qualitatively different behavior.  Fig. \ref{Fig:ScaleNode} 
presents $A(\bk,\omega)$ along the nodal cut ((0,0) - ($\pi$,$\pi$)) for the same three samples.  The 
highlighted region indicates the approximate position of the kink.  
In the nodal region the overall bandwidth is much larger than the energy of the bosonic modes so  
the dramatic band breakup does not occur \cite{tpdPRL2004} and the renormalization manifests as a kink 
in the dispersion.  In this case, the energy scale of the kink is most easily determined from  
the structure of the real part of the self-energy Re$\Sigma$.  The MDC-derived estimate for 
Re$\Sigma$, obtained from subtracting the MDC-derived dispersion from an assumed 
linear band, is also in the lower panels of Fig. \ref{Fig:ScaleNode}.  

\begin{figure}[t]
 \vskip 0.5cm
 \includegraphics[width=\columnwidth]{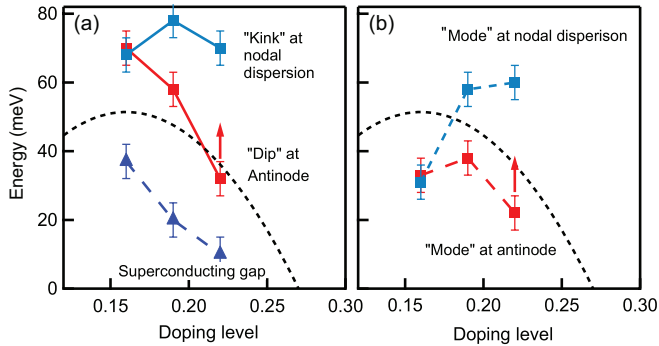}
 \caption{\label{Fig:Summary} A summary of some energy scales relevant to the renormalized band 
 dispersions. (a) The apparent kink position in the nodal dispersion, dip energy at the antinodal region 
 and the superconducting gap are summarized for the three doping levels shown in Fig 
 \ref{Fig:ScaleAntinode}. (b) the mode energy obtained by subtracting the superconducting gap from the 
 characteristic energies of the renormalization effect.  The red arrow is to remind the 
 reader that the shown quantity at the antinodal region of the OD65K sample is a lower bound for 
 the actual value.}
\end{figure}

The nominal doping dependence \cite{Perslan91} of the energy scales in the nodal and antinodal region 
of Bi2212 are summarized in Fig. \ref{Fig:Summary}a.  For reference, 
the superconducting gap $\Delta_0$ is shown also, which is determined from the peak
positions of the Fermi function divided spectrum at the Fermi level.  
While the characteristic energy in the antinodal region (dip energy) 
follows the decrease in the superconducting gap, the characteristic energy in the nodal region 
remains more or less constant ($\sim 70$ meV).  The difference in the doping 
dependence of the two energy scales lends further support to the existence of coupling to multiple 
modes.  If a single mode were responsible for the renormalization throughout the zone one would 
expect the doping dependence to follow the same trend in the nodal and antinodal regions.   

The energy of the dominant mode $\Omega$ can be obtained by subtracting the magnitude of the 
superconducting gap from the observed energy scale, expected to be $\Omega + \Delta_0$\cite{tpdPRL2004}. 
The results of this procedure are shown in Fig. \ref{Fig:Summary}b.  The energy of the dominant mode 
in the antinodal region, within the error bars of the data, is independent of doping.  
The behavior in the nodal region is different; the energy of the dominant mode changes with doping.  
At optimal doping the energy of the dominant mode is $\sim 35$ meV but in 
the overdoped samples the energy is larger $\sim 60$ meV.  
This result is consistent with the picture of coupling to multiple modes outlined in 
Ref. \onlinecite{tpdPRL2004}.  We also note that in OD88K a secondary feature can be observed 
in Re$\Sigma$ at precisely the same energy as the dip energy of the antinodal region.  
Similar fine structure was reported earlier in Ref. \onlinecite{LeePRB2008}.  The presence of this 
sub-feature in the UD88K data as the sample is progressively overdoped, along with the 
35 meV scale in the nodal data at optimal doping,  
is evidence of a trade off between a coupling dominated by the B$_{1g}$ mode and 
one dominated by the bond stretching mode.  We further note that Ref. \onlinecite{GromkoPRB2003} 
reached similar conclusions but assigned the anti-nodal renormalizations to the spin resonance mode.  
We believe that the multi-layer data of the previous section, especially the single-layer Tl data which 
shows no renormalization in the anti-nodal region, directly refutes this conclusion and favors the el-ph 
scenario.  

\section{Self-doping}
We now turn our attention to the identity of the carrier types in the inner and outer planes 
in F0234.  Experimentally, the parent compound of F0234 is known to self-dope with the 
inner and outer planes having Fermi surfaces of different carrier type.   
A recent ARPES study\cite{ChenPRL2009} found that the $p$-type 
bands are bilayer split along the nodal direction.  Since the inner planes are expected to have a 
stronger inter-planar coupling than the outer planes, the observation of bilayer splitting 
provides strong evidence that the inner layers are $p$-type while the outer layers are $n$-type. 
Furthermore, kinks in the nodal dispersion of both sets of planes but the strength of the 
coupling in the $n$-type layer by a factor of two.    

\begin{figure}
 \includegraphics[width=\columnwidth]{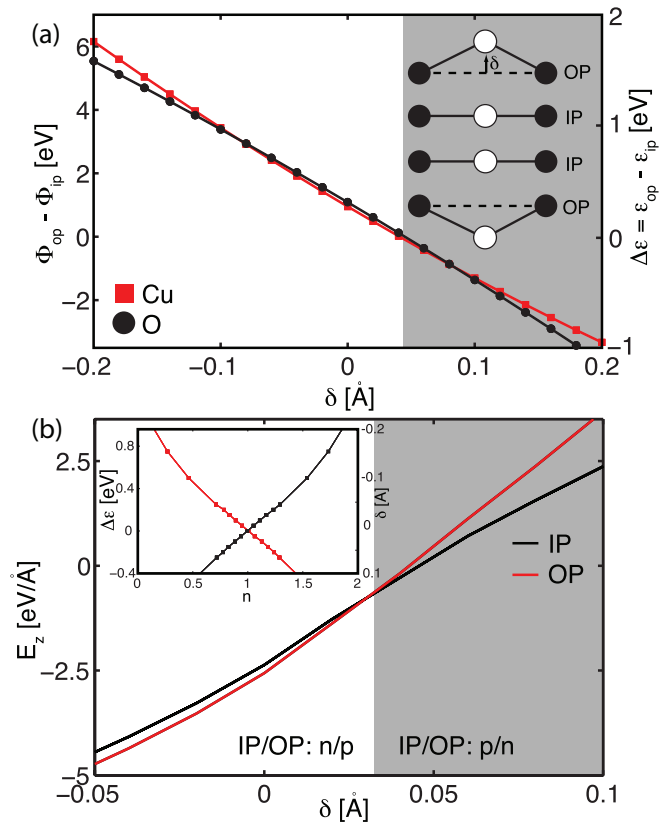}
 \caption{\label{Fig:Madelung}
 A summary of the electrostatic calculations for F0234.  (a) The Madelung potential  
 difference between the Cu (red) and O (black) sites of the inner and outer planes as a function 
 of the oxygen buckling of the outer layer.  The shaded region indicates the region where the 
 inner and outer planes are identified as $p$- and $n$-type respectively.  Buckling of the planar 
 O away from the mirror plane of the crystal is defined to be positive.  (b, inset) The 
 filling of the inner (red) and outer (black) planes as a function of the site energy difference 
 (buckling) between inner and outer layers.  Fillings are obtained with a simple tight-binding model 
 (see text).  (b) The electric field at the oxygen site of the inner (black) and outer (red) 
 layers as a function of the buckling distance.  
 }
\end{figure}

The doping of individual layers will be driven by Madelung energy differences between the 
inner and outer planes. 
In order to determine the Madelung potential $\Phi$ of each site we again employ the Ewald summation 
technique using the structural data from Ref. \onlinecite{Structure} and assigning formal valence 
charges to each atom.  In principle, both the inner and outer Cu sites, as well as the inner and outer O sites,  
can have different Madelung energies.  The in-plane difference  
$\Delta\Phi^\alpha = \Phi^\alpha_{O} - \Phi^\alpha_{Cu}$, where $\alpha$ is a plane index,  determines 
the charge transfer energy\cite{JohnstonEPL2009} $\Delta^\alpha$ for each plane.  The difference between layers 
$\Delta\Phi^\perp_i = \Phi_{i}^{op} - \Phi_i^{ip}$, where $i = $ Cu or O, controls the relative 
site energies of the two planes.  However, our Ewald calculations show that $\Delta^{ip} \sim \Delta^{op}$ 
and $\Delta\Phi^\perp_{Cu} \sim \Delta\Phi^\perp_O$.  We therefore neglect 
the differences and take the charge transfer energy to be the same in each layer but shift the outer 
plane's site energies by $\Delta\epsilon^\perp = \Delta\Phi/\epsilon(\infty)$ where  
$\epsilon(\infty)$ is the dielectric constant which taken to be $3.5$. 
 
To understand the self-doping phenomena we consider a simple tight-binding model for the CuO$_2$ planes.  
For simplicity, the form for the bandstructure is taken from a 5-parameter tightbinding fit to the 
low-energy dispersion of Bi2212\cite{NormanPRB1995} as a representative singleband model.     
We also introduce the usual inter-planar coupling term\cite{XiangPRL1996}  
$\epsilon_\perp(\bk) = t_{perp}(\cos(k_xa) - \cos(k_ya))^2/4$ with $t_\perp = 50$ meV.  
The resulting model Hamiltonian is

\begin{eqnarray}\label{Eq:hamiltonian}\nonumber
 H&= \sum_{\alpha=1}^4\sum_{\bk,\sigma} (\epsilon_{\alpha}(\bk) - \mu)d^\dagger_{\alpha,\sigma,\bk}d_{\alpha,\sigma,\bk} \\ 
 &+\sum_{<\alpha,\alpha^\prime>} \epsilon_\perp(\bk)[d^\dagger_{\alpha,\sigma,\bk}d_{\alpha^\prime,\sigma,\bk} + H.C.]
\end{eqnarray}

where $\alpha = 1-4$ is the plane index, $\epsilon_\alpha(\bk) = \epsilon(\bk) + \Delta\Phi^\perp/\epsilon(\infty)$ 
for the outer planes and $\epsilon_\alpha(\bk) = \epsilon(\bk)$ for the inner planes 
and $<...>$ is a sum over neighboring planes.  (The in-plane charge transfer energy has been absorbed into the 
definition of $\epsilon(\bk)$.) Finally, $\mu$ is the 
chemical potential which is adjusted to maintain the total filling of the parent compound, 
$\sum_{\alpha,\sigma} \hat{n}_{\alpha,\sigma} = 4$.  
The resulting model is then diagonalized for a given $\Delta\Phi$ in order to obtain the relative filling 
of the four planes.  

In previous LDA treatments of this material\cite{XiePRL2007}, it was found that 
the identification of the carrier types in each layer is dependent on the degree of Cu-O buckling 
occurring in the outer layers.  We therefore calculate the Madelung potential difference between the 
inner and outer layers as a function of the degree of buckling.  The results are shown in Fig. 
\ref{Fig:Madelung}a.  For flat or dimpled ($\delta < 0$) planes, $\Delta\Phi = \Phi_{op} - \Phi_{ip}$ 
is positive indicating a larger site energy in the outer layer.  Charge is therefore expected to flow 
from the outer planes (OP) to the inner planes (IP) and the OP/IP are identified as $p$- and $n$-type respectively.  
As the plane is buckled outward $\Delta\Phi$ is suppressed and becomes 
negative for $\delta \sim 0.04$ $\AA$ at which point the identity of the planes is reversed.  The filling of 
the IP and OP, obtained from diagonalizing (\ref{Eq:hamiltonian}), is shown in the inset of 
Fig. \ref{Fig:Madelung}b.  

Once the fillings have been obtained we uniformly transfer charge between the 
planes and recalculate the crystal field at the oxygen sites. The resulting fields are plotted in Fig.  
\ref{Fig:Madelung}b.   For a dimpled plane the electric field at the IP and OP roughly follow a linear 
dependence on the buckling distance with the IP having only a slightly weaker field strength.  In this case, the 
kink strength measured by ARPES in the two bands would be comparable.  For a buckled plane, once the sign of the 
Madelung potential difference changes and the identification of the planes reverses, deviations in the field 
strength become more pronounced.  Since the coupling enters as $E^2$ the $n$-type outer plane is expected 
to have a stronger kink feature, consistent with photoemission experiments.  

The carrier concentration of the two sets of planes, determined from the Luttinger fraction, were reported in Ref.  
\onlinecite{ChenPRL2006} with dopings of $0.60 \pm 0.04$ and $0.4 \pm 0.03$ in the $p$- and $n$-type 
bands.  This observation, coupled with the different kink strength in each of the bands corresponds to  
$\delta = 0.1$ $\AA$ in our calculations.  For this degree of buckling, the ratio of the crystal fields is 
$E_{op}/E_{ip} = 1.55$ and the ratio of the kink strength is $\lambda_{op}/\lambda_{ip} = 2.4$.  
(The absolute value of $\lambda$ depends on the form of the coupling constant $g(\bk,\bq)$\cite{tpdPRL2004,LeePRB2007}.) 
This ratio is slightly larger than the ratio of $\sim 2$ reported in Ref. \onlinecite{ChenPRL2009} however the agreement 
is good considering the simplicity of the model.  It is also clear that the inequivalent dopings can be understood 
from electrostatic configurations.  Furthermore, the asymmetry in the coupling in each layer falls naturally into 
the $c$-axis phonon scenario with the strength of the coupling being further driven by the self-doping process.   

\section{Conclusions}
In this work we have presented aspects of the material and doping dependence of the dispersion 
renormalizations in the nodal and antinodal regions of various single- and multi-layer cuprates.  
We have found that the strength of the nodal kink has a strong material dependence and varies with 
the number of layers present in the material.  In general, the kink strength mirrors T$_c$, 
taking on a maximal value in the $n = 3$ compounds.  The issue can be complicated further  
in the multi-layer cuprates, where Madelung potential differences can lead to inequivalant dopings 
in the various layers.  This can lead to further symmetry breaking 
across the CuO$_2$ planes, and results in different kink strength in the different layers within the 
same material.  Using a simple tight-binding model and electrostatic calculations we have developed 
a picture of this phenomena in self-doped F0234, which is consistent with recent ARPES studies.    

The renormalization in the antinodal region also shows 
a marked dependence on the number of layers present in the material and is unresolved in the 
single layer cuprates.  This result is difficult to reconcile for coupling to the spin resonance 
mode, which is expected not to vary with the number of layers, but is naturally explained by 
coupling to the B$_{1g}$ phonon when one considers the crystal structure of these materials.  

Further evidence for multiple phonon modes was found in the doping dependence of Bi-2212.  Here, 
the features in the nodal and antinodal regions exhibit different behavior.  Once gap referenced, 
the energy scale in the nodal region changes from $\sim 35-40$ meV to $\sim 70-80$ meV as the sample 
is overdoped. This change of energy scales cannot be explained by coupling to a single mode and therefore rules 
out the spin resonance mode, at least as the sole player.   
In the phonon scenario this signifies a trade-off between dominant coupling to the B$_{1g}$ mode near optimal 
doping and a dominant coupling to the bond stretching mode in the overdoped samples.  The change of relative coupling 
is due to increased screening of the B$_{1g}$ mode as the carrier concentration is increased.  

Both the doping and materials dependence presented here provides compelling evidence 
that a spectrum of phonon modes are responsible for both the nodal and antionodal low-energy renormalizations 
observed in the cuprates.  

\section{Acknowledgements}
This work was supported by the US Department of Energy, Office of Basic Energy Science under 
contracts DE-AC02-76SF00515 and DE-AC02-05CH11231.  
Portions of this research were carried out at the Stanford Synchrotron Radiation Lightsource (SSRL), 
a national user facility operated by Stanford University on behalf of the US Department of Energy, 
Office of Basic Energy Sciences. 
This research used resources of the National Energy Research Scientific Computing Center, 
which is supported by the Office of Science of the U.S. Department of Energy under Contract No. DE-AC02-05CH11231.  
S. J. would like to acknowledge support from NSERC and SHARCNET.


\begin{thebibliography}{99}
\bibitem{Perslan91}
The nominal doping level is estimated from T$_c$ using the empirical law 
$T_c = T_{c}^{\rm{max}}[1-82.6(x-0.16)^2]$.
\end{thebibliography}
\end{document}